\documentclass[%
 reprint,
nofootinbib,
 amsmath,amssymb,
 aps,
pra,
]{revtex4-2}

\usepackage{titlesec}
\titlespacing*{\section}{0pt}{1.1\baselineskip}{\baselineskip}
\titlespacing*{\subsection}{0pt}{1.1\baselineskip}{\baselineskip}

\usepackage[utf8]{inputenc}
\usepackage{braket}

\usepackage{graphicx}
\usepackage{dcolumn}
\usepackage{bm}
\usepackage{hyperref}

\usepackage{mathrsfs}

\usepackage{relsize}

\usepackage{listings}

\begin{document}

\title{Parallelising the Queries in Bucket Brigade Quantum RAM}

\author{Alexandru Paler$^{+,*}$, Oumarou Oumarou$^{\pi}$, Robert Basmadjian$^{\pi}$}
\affiliation{
    $^{+}$University of Transilvania, B-dul Eroilor 29, 500036, Brașov, România
}
\affiliation{
    $^{*}$Johannes Kepler University, Altenberger Str. 69, 4040, Linz, Austria
}
\affiliation{
    $^{\pi}$Department of Informatics, Clausthal University of Technology, 38678 Clausthal-Zellerfeld, Germany
}

\email{alexandrupaler@gmail.com}

\begin{abstract}
Quantum algorithms often use quantum RAMs (QRAM) for accessing information stored in a database-like manner. QRAMs have to be fast, resource efficient and fault-tolerant. The latter is often influenced by access speeds, because shorter times introduce less exposure of the stored information to noise. The total execution time of an algorithm depends on the QRAM access time which includes: 1) address translation time, and 2) effective query time. The bucket brigade QRAMs were proposed to achieve faster addressing at the cost of exponentially many ancillae. We illustrate a systematic method to significantly reduce the effective query time by using Clifford+T gate parallelism. The method does not introduce any ancillae qubits. Our parallelisation method is compatible with the surface code quantum error correction. We show that parallelisation is a result of advantageous Toffoli gate decomposition in terms of Clifford+T gates, and after addresses have been translated, we achieve theoretical $\mathcal{O}(1)$ parallelism for the effective queries. We conclude that, in theory: 1) fault-tolerant bucket brigade quantum RAM queries can be performed approximately with the speed of classical RAM; 2) the exponentially many ancillae from the bucket brigade addressing scheme are a trade-off cost for achieving exponential query speedup compared to quantum read-only memories whose queries are sequential by design. The methods to compile, parallelise and analyse the presented QRAM circuits were implemented in software which is available online.
\end{abstract}

\maketitle

\section{Introduction}
\label{sec:intro}

Quantum random access memory (QRAM) was proposed for querying (e.g. reading and writing) information during quantum information processing. However, their practical utility is still under study due to the exponential costs associated to explicitly storing and retrieving information \cite{schuld2018supervised}. Nevertheless, QRAM and their read-only-variant QROM (quantum read-only memories) \cite{babbush2018encoding} are still capturing the attention of researchers due to their conceptual similarity to classical RAMs, and due to the fact that they can be used in subroutines of relevant quantum algorithms, such as the one for linear systems of equations \cite{coles2018quantum}.

Whenever QRAM queries are frequent, the speed of the QRAMs influences the total execution time of the quantum algorithms. Moreover, the total execution time of an algorithm impacts the amount of resources necessary for achieving fault-tolerance \cite{paler2019really}. Fast QRAMs may prove useful in implementing large scale fault-tolerant quantum computations. Implementations of QRAM circuits were presented in \cite{oum2020} and available online\footnote{\url{https://github.com/quantumresource/quantify}}.

The access time of a QRAM includes: 1) the time necessary to determine, based on an input set of addresses, the memory locations to query, and 2) the time required to perform the effective queries (e.g. retrieve or store information). The bucket brigade QRAM variant \cite{arunachalam2015robustness} was proposed to reduce the address translation time necessary for computing the memory pointers where the queries should be executed.

\begin{figure}[t!]
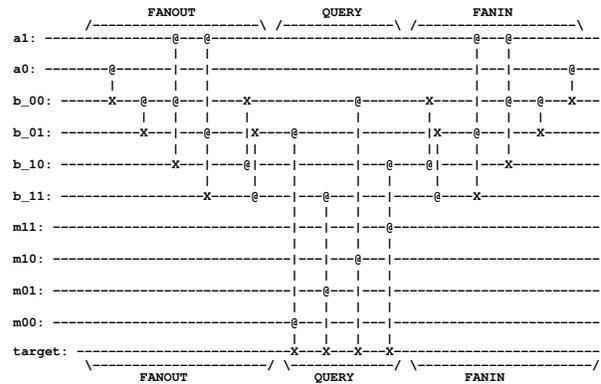

\linespread{.5}
\begin{lstlisting}[basicstyle=\ttfamily\bfseries\tiny]
                 FANOUT                QUERY              FANIN
         /---------------------\ /--------------\ /--------------------\
a1: ----------------@---@---------------------------------@---@-----------
                    |   |                                 |   |
a0: --------@-------|---|---------------------------------|---|-------@---
            |       |   |                                 |   |       |
b_00: ------X---@---@---|----X-------------@--------X-----|---@---@---X---
                |   |   |    |             |        |     |   |   |
b_01: ----------X---|---@----|X----@-------|--------|X----@---|---X-------
                    |   |    ||    |       |        ||    |   |
b_10: --------------X---|----@|----|-------|---@----@|----|---X-----------
                        |     |    |       |   |     |    |
b_11: ------------------X-----@----|---@---|---|-----@----X---------------
                                   |   |   |   |
m11: ------------------------------|---|---|---@--------------------------
                                   |   |   |   |
m10: ------------------------------|---|---@---|--------------------------
                                   |   |   |   |
m01: ------------------------------|---@---|---|--------------------------
                                   |   |   |   |
m00: ------------------------------@---|---|---|--------------------------
                                   |   |   |   |
target: ---------------------------X---X---X---X--------------------------
         \----------------------/ \-------------/ \----------------------/
                FANOUT                QUERY              FANIN
\end{lstlisting}
\caption{A bucket brigade circuit consisting of CNOTs and Toffoli gates. The controlled gates from the diagram use @ for controls and the X for target wires. The address wires are named $a_i$, the memory cells holding single bits are $m_i$. The classical bits stored on the wires $b_i$ determine which memory wires to read/write.}
\label{fig:buck2}
\end{figure}

QRAMs can be described in the form of quantum circuits. Information is stored to and retrieved from the wires/qubits (e.g. analogue to classical memory cell), while quantum gates are used to implement address translation and the queries (e.g. effective read and write operations). In this work, we focus on retrieving information from the bucket brigade QRAM because, due to their structure, reading is more expensive than writing. Once the addresses have been translated, the bucket brigade requires Toffoli (double-controlled-X, or CCX) gates for reading, but only CNOT (controlled-NOT, or controlled-X) gates for writing. We assume the information is already stored in the QRAM database, and that the quantum algorithm queries frequently the QRAM.

For a QRAM with $2^q$ distinct memory wires (and exponential number of ancillae wires, each representing a memory cell that stores a single bit, such as in \cite{di2019fault}), there are $q$ bits necessary to address any of the wires. A QRAM circuit takes as input a superposition $\ket{adr_j}$ addresses to be queried ($\alpha_j$s are complex amplitudes), and outputs the contents of the memory cell $m_j$ addressed by $\ket{adr_j}$. 

\begin{align*}
\sum_{j=0}^{2^q-1}\alpha_j\ket{adr_j}\ket{0} \xrightarrow{QRAM} \sum_{j=0}^{2^q-1}\alpha_j\ket{adr_j}\ket{m_j}
\end{align*}

A bucket brigade QRAM circuit consists of three sub-circuits (regions): 1) FANOUT where an exponential number of ancilla bits $b_i$ is used to store control bits for pointing to the corresponding memory cells $m_i$; 2) QUERY where the memory bits $m_i$ are queried and the results are stored on the ancilla wire $\ket{target}$; 3) FANIN (the uncomputation of FANOUT) in order to maintain reversibility, where the $b_i$ bits are returned to their initial $\ket{0}$ state. The three sub-circuits can be easily recognised in Fig.~\ref{fig:buck2}. The QUERY region includes only Toffoli gates.

From the quantum circuit perspective: 1) address-translation duration is equivalent to the circuit depth of the FANOUT/FANIN regions; 2) the effective query speed is analogous to the depth of the QUERY region. The trade-off cost for speeding-up of the address translation is the exponentially large number of $b_i$ ancillae, which translates to additional resource requirement and hence increase in circuit's width. In the standard bucket brigade QRAM model, the queries are considered strictly sequential. Thus, the depth of the QUERY region is dictated by the number of queries. In the worst case, where the entire QRAM is queried, the depth of QUERY is exponentially long \cite{bang2019optimal, di2019fault}.

The state-of-the-art Clifford+T formulation of the bucket brigade quantum circuits \cite{di2019fault} was obtained from the original formulation \cite{arunachalam2015robustness} and decomposing the Toffoli gates into known Clifford+T gate sequences. The authors of \cite{di2019fault} mentioned that important resource savings could be possible if the CCZ (double-controlled-Z gate)/Toffoli transformation is employed in an advantageous manner for the compilation of QRAM. The CCZ gate is obtained from a Toffoli and two Hadamard gates.

This work continues on that line of thought and illustrates some of the possible improvements. First, the depth of the QUERY region is exponentially reduced by parallelising the Toffolis. Second, the herein presented method achieves a linear width improvement compared to the state-of-the-art from \cite{di2019fault}, where parallelisation was obtained by introducing four ancillae for each Toffoli gate. We show that our method is ancillae free, and reduces the width by a factor of four. Third, the presented QRAMs have a T-count which is almost double to QROM \cite{babbush2018encoding}, and approximately half compared to the QRAMs from \cite{di2019fault} (c.f. $T_{bbs}$, $T_{rom}$, $T_{bbp}$ in Sec.~\ref{sec:theoretical_speedup}).

\begin{figure}[t!]
    \setlength{\abovecaptionskip}{1mm}
    \centering
    \includegraphics[width=0.9\columnwidth]{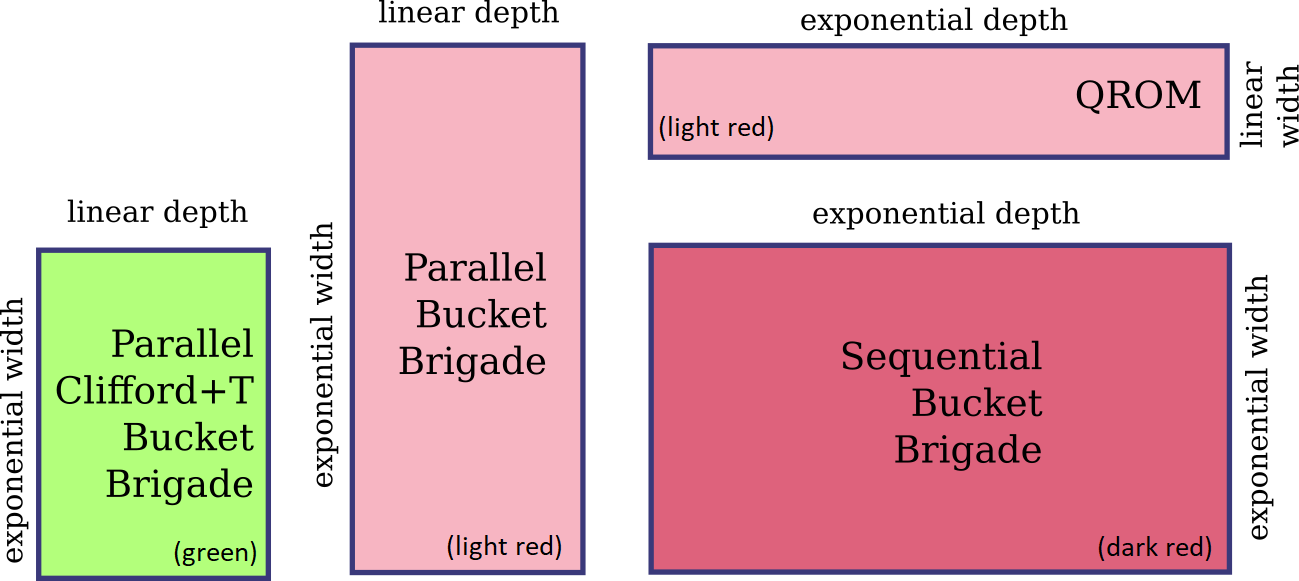}
    \caption{Comparison between different versions of QRAM. The proposed QRAM is marked green. The parallel bucket brigade (light red) \cite{di2019fault} introduces an exponential number of ancillae to achieve parallelism. This is not the case with our decomposition. The original (sequential) bucket brigade circuit (dark red) is sequential and has the same number of wires like the one proposed herein, but requires an exponential number of control signals. The QROM \cite{babbush2018encoding} circuit was specifically designed with sequential queries, but does not require the control signals.}
    \label{fig:compare}
\end{figure}

The Toffoli construction used in this paper can be interpreted as the enabler of trading depth for width between QROM and QRAM: it exponentially increases the width of the QRAM circuit, while it exponentially decreases the depth of the circuit. Fig.~\ref{fig:compare} illustrates the improvements obtained by parallelising the QRAM.

In the following, the Results section presents exact resource counts for the proposed bucket brigade QRAMs, and includes a detailed comparison to other QRAM variant layouts. The Methods section discusses how quantum gate level parallelism is achieved, and illustrates how Clifford+T decompositions of Toffoli gates can be used in an advantageous manner for each QRAM sub-circuit. Finally, we present empiric evidence to conjecture that the presented optimisation technique could be applicable to many types of circuits based on multi-controlled-operations \cite{barenco1995elementary}, which are the common building blocks of quantum adders and multipliers.

In order to achieve a fair comparison between QRAM circuit implementations, we consider $n \leq q$, with $2^n$ being the number of memory queries, and $2^q$ denoting the number of memory cells $m_i$, such that $0 \leq i<2^q$. The worst case for the number of queries, as used in \cite{di2019fault}, is for $n=q$.

\section{Results}
\label{sec:theoretical_speedup}

The advantage of bucket brigade QRAM queries is that these can be parallelised. The presented method is implemented in an open-source software\footnote{\url{https://github.com/quantumresource/quantify}}\cite{oum2020}.

The gate level parallelism used in this work is conceptually very similar to classical instruction level parallelism (see Sec.~\ref{sec:par}). In practice, the QRAM gate parallelism was lost when translating the circuits to Clifford+T -- an often necessary step for estimating the computational resources required for quantum error-correction \cite{paler2019really}. The width (number of wires, not including the exponential number of memory cells $m_i$), T-count, and the depth of the QRAM circuits \cite{di2019fault} are:
\begin{align*}
    Q_{bbs} & = q + 2^q + 5\\
    T_{bbs} & =  21\cdot2^q - 28\\
    D_{bbs} & =  21\cdot2^q + 2q - 26
\end{align*}

Since their initial formulation, a disadvantage of bucket brigade QRAMs is the fact that an exponentially high number of ancilla $b_i$ bits has to be computed in order to achieve FANOUT parallelism. The exponential overhead is far from ideal, when considering that quantum hardware resources are and will remain very scarce. For this reason, the QROM was proposed in \cite{babbush2018encoding}. It does not require the $b_i$ bits, but queries are strictly sequentialised (one after the other). By using optimised Toffoli decompositions, the T-count of the QROM is linear in the number of queries $2^n$ (the authors of \cite{babbush2018encoding} used $L$ for the number of queries). The formulas for width ($Q_{rom}$) and depth ($D_{rom}$) do not refer to the memory cells $m_i$.
\begin{align*}
Q_{rom} & = q + 1\\
T_{rom} & =  4\cdot2^n - 4\\
D_{rom} & =  10 \cdot 2^n + c\cdot\mathcal{O}(2^n)
\end{align*}

We present a bucket brigade construction that has a reduced T-count, and a depth exponentially shallower than the circuits from \cite{di2019fault}. Our construction is achieved using an advantageous parallelisable CCZ/Toffoli decomposition.

\begin{align}
    Q_{bbp} & = q + 2^q + 1\label{eq:q}\\
    T_{bbp} & =  T_{fanout} + T_{query} = (4 \cdot 2^q) + (6 \cdot 2^n)\label{eq:t}\\
    D_{bbp} & =  D_{fanout} + D_{query} + D_{fanin} = (10 \cdot q) + 10 + (4 \cdot q)\label{eq:d}
\end{align}

Our construction has a $Q_{bbp}$ which is the same as the one from \cite{di2019fault, arunachalam2015robustness}. We do not modify the width of the QRAM, but we optimise the depth of the QRAM. The equivalent quantum volume of the circuit being protected by the surface code is not evaluated in this work, although the parallelism is achieved by using a surface-code compliant parallel CNOT construction. The hardware footprint of bucket brigade QRAM is prohibitively high -- this is dictated mostly by the width of the circuit. Having exponentially reduced the QRAM depth and the T-count, maintains the exponentially large width (compared to QROM, $Q_{rom} \lll Q_{bpp}$). For future work, it will be reasonable to estimate the resources necessary for surface code protection, especially when these QRAMs are included in practical quantum algorithms.

\begin{figure}
    \setlength{\abovecaptionskip}{1mm}
    \centering
    \includegraphics[width=0.8\columnwidth]{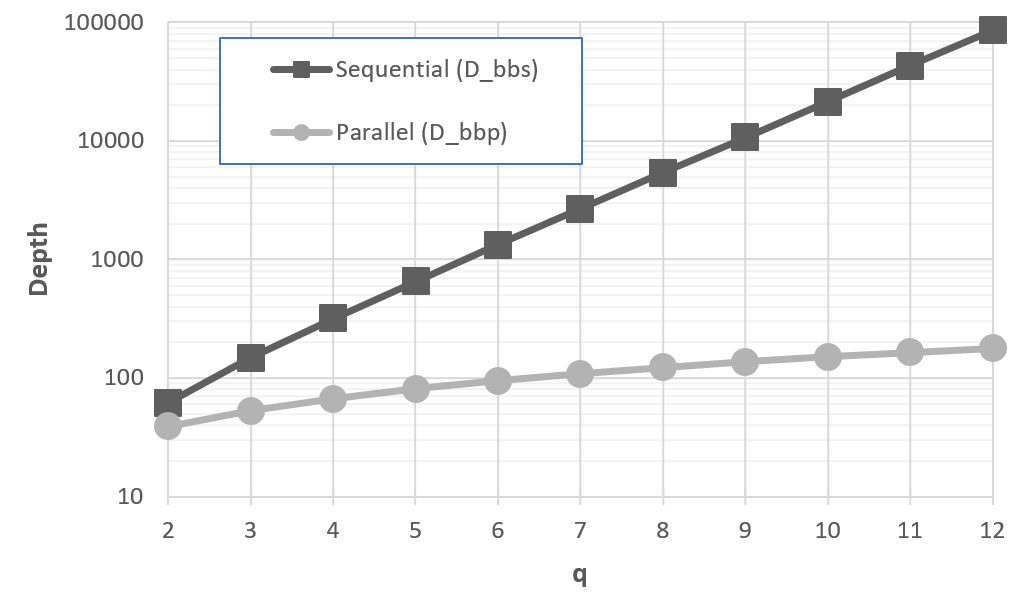}
    \caption{Comparison between $D_{bbs}$ and $D_{bbp}$. This plot assumes the worst case where the number of queries $2^n$ equals the number of memory cells $2^q$. The exponent $q$ is plotted along the horizontal axis.}
    \label{fig:depth}
\end{figure}

\begin{figure}[h!]
    \setlength{\abovecaptionskip}{1mm}
    \centering
    \includegraphics[width=0.7\columnwidth]{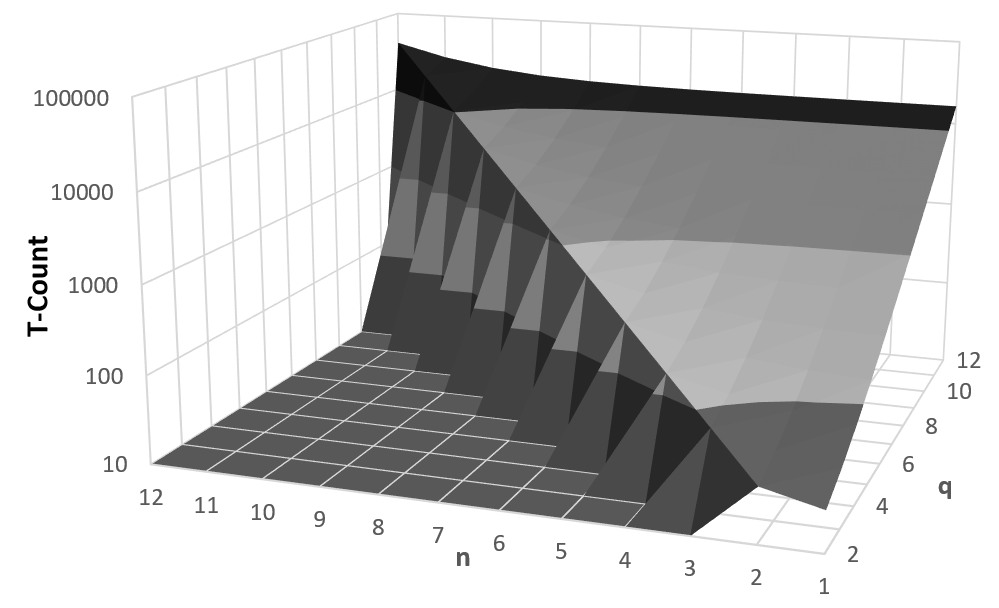}
    \caption{The T-count of $T_{bbp}$. The number of queries ($2^n$) is varied by the parameter $n$, and the number of memory cells $2^q$ by the parameter $q$, such that $n\leq q$. The T-count increases exponentially with $q$, and the depth (cf. Fig.~\ref{fig:depth}) is linear in $q$.}
    \label{fig:tcount}
\end{figure}

\begin{figure*}
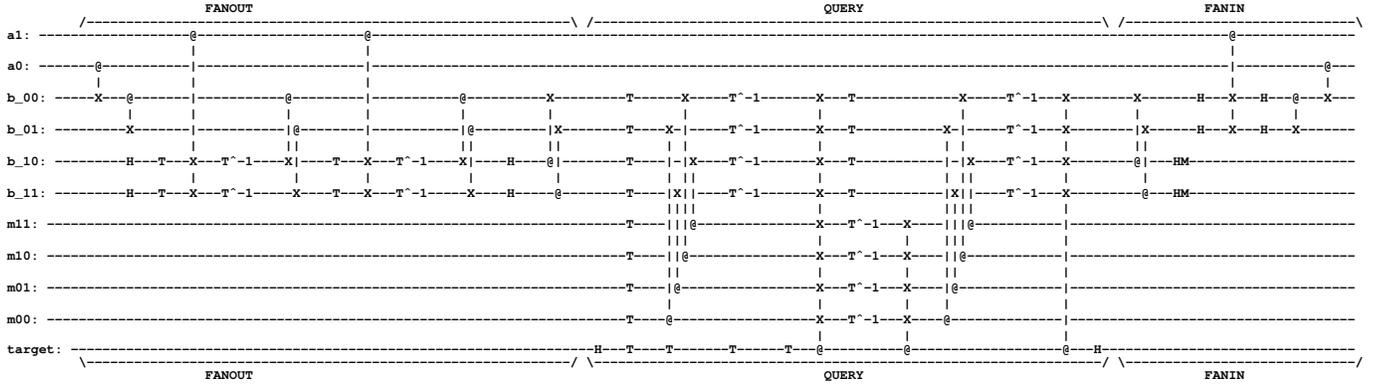

\linespread{.5}
\begin{lstlisting}[basicstyle=\ttfamily\bfseries\tiny]
                         FANOUT                                                                        QUERY                                           FANIN
         /-------------------------------------------------------------\ /----------------------------------------------------------------\ /-----------------------------\
a1: -------------------@---------------------@------------------------------------------------------------------------------------------------------------@---------------
                       |                     |                                                                                                            |
a0: -------@-----------|---------------------|------------------------------------------------------------------------------------------------------------|-----------@---
           |           |                     |                                                                                                            |           |
b_00: -----X---@-------|-----------@---------|-----------@----------X---------T------X-----T^-1-------X---T-------------X-----T^-1---X--------X-------H---X---H---@---X---
               |       |           |         |           |          |                |                |                 |            |        |           |       |
b_01: ---------X-------|-----------|@--------|-----------|@---------|X--------T----X-|-----T^-1-------X---T-----------X-|-----T^-1---X--------|X------H---X---H---X-------
                       |           ||        |           ||         ||             | |                |               | |            |        ||
b_10: ---------H---T---X---T^-1----X|----T---X---T^-1----X|----H----@|--------T----|-|X----T^-1-------X---T-----------|-|X----T^-1---X--------@|---HM---------------------
                       |            |        |            |          |             | ||               |               | ||           |         |
b_11: ---------H---T---X---T^-1-----X----T---X---T^-1-----X----H-----@--------T----|X||----T^-1-------X---T-----------|X||----T^-1---X---------@---HM---------------------
                                                                                   ||||               |               ||||           |
m11: -------------------------------------------------------------------------T----|||@---------------X---T^-1---X----|||@-----------|------------------------------------
                                                                                   |||                |          |    |||            |
m10: -------------------------------------------------------------------------T----||@----------------X---T^-1---X----||@------------|------------------------------------
                                                                                   ||                 |          |    ||             |
m01: -------------------------------------------------------------------------T----|@-----------------X---T^-1---X----|@-------------|------------------------------------
                                                                                   |                  |          |    |              |
m00: -------------------------------------------------------------------------T----@------------------X---T^-1---X----@--------------|------------------------------------
                                                                                                      |          |                   |
target: ------------------------------------------------------------------H---T----T-------T------T---@----------@-------------------@---H--------------------------------
         \-------------------------------------------------------------/ \----------------------------------------------------------------/ \-----------------------------/
                         FANOUT                                                                        QUERY                                           FANIN
\end{lstlisting}
\caption{The equivalent Clifford+T representation of the circuit from Fig.~\ref{fig:buck2}. The depth of the circuit is constant for a fixed $q$ by FANIN and FANOUT, while QUERY has a constant depth irrespective of $q$ or $n$. The circuit operates on $2^q$ memory cells and executes $2^n$ queries with $n \leq q$. The sequence of four T gates on the target qubit will be cancelled.}
\label{fig:complete}
\end{figure*}

\section{Methods}

This work assumes that, in general, two gates $G_1$ and $G_2$ can be parallelised if their commutator $[G_1,G_2] = 0$. From a functional perspective, it does not matter in which order the gates are applied: in an ideal setting, the gates can be executed without one depending on the output of the other. This approach to quantum gate parallelism is very similar to classical computing instruction-level parallelism.

\subsection{Parallel CNOTs}
\label{sec:par}

\begin{figure}[h]
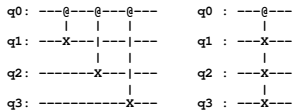

\linespread{.5}
\begin{lstlisting}[basicstyle=\ttfamily\bfseries\tiny]
                  q0: ---@---@---@---     q0 : ---@---
                         |   |   |                |
                  q1: ---X---|---|---     q1 : ---X---
                             |   |                |
                  q2: -------X---|---     q2 : ---X---
                                 |                |
                  q3: -----------X---     q3 : ---X---
\end{lstlisting}
\caption{Parallel CNOTs are allowed to share the control wire.}
\label{fig:parcnot}
\end{figure}

For the purpose of this work, two CNOTs are parallel if they either share the controls, or no wires at all (see Fig. \ref{fig:parcnot}). CNOTs can also share the target, but this fact will not be used herein. Most of the Clifford+T decompositions are compiled and optimised due to the inherent necessity for quantum error-correction, and this kind of CNOT parallelism is supported by the surface code \cite{paler2017synthesis} (braided and lattice surgery variants), for example.

\subsection{Parallelisable CCZ}

The three qubit Toffoli gate is in general derived from the CCZ gate, which is a controlled application of the two-qubit CZ gate. The CCZ flips the sign of phase of a state if there is a $\ket{1}$ on all of the gate's three input wires (qubit and wire are used interchangeably). Thus, the phase flip can be expressed as $(-1)^{xyz}$, where $x,y,z$ are the bit values of the inputs: the phase is multiplied by $-1$ if all three bits are $1$, and no phase flip is applied otherwise ($(-1)^0 = 1$). It has been shown by \cite{barenco1995elementary}, and more recently in \cite{gidney2018halving}, that the following Boolean formula is useful for expressing the CCZ through CNOTs and T gates.
\begin{align}
\label{eq:1}
4xyz = x + y + z - (x \oplus y) - (y\oplus z) - (x\oplus z) +(x\oplus y \oplus z)
\end{align}

The T gate rotates the phase of a state by $\frac{\pi}{4}$, and $-1=\omega^4=e^{i\frac{\pi}{4}}$, such that $(-1)^{xyz}=\omega^{4xyz}$. Thus, seven T gates are necessary, each conditioned on one of the parity sums from Eq.~\ref{eq:1}.

Eq.~\ref{eq:1} is a recipe for generating valid CCZ gate decompositions. Seven parity sums are computed by using CNOTs and T gates, while ensuring that two conditions are met. The first condition is for the T gates to be applied at the right moment, when the necessary bit parities are stored on any of the wires. Second, the parities on each of the three wires have to be uncomputed in order to reflect a correct functionality. Ancillae used for parity computations have to be uncomputed, too. Due to the form of Eq. \ref{eq:1}, containing seven parities, the number of ancillae seems to be bounded by seven, but in practice the maximum is four, because three of the parities are formed by single bit values. Consequently, the literature includes a large number of decompositions of the CCZ in terms of Clifford+T gates.

\begin{figure}[h!]
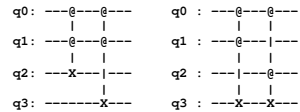

\linespread{.5}
\begin{lstlisting}[basicstyle=\ttfamily\bfseries\tiny]
                  q0: ---@---@---     q0 : ---@---@---
                         |   |                |   |
                  q1: ---@---@---     q1 : ---@---|---
                         |   |                |   |
                  q2: ---X---|---     q2 : ---|---@---
                             |                |   |
                  q3: -------X---     q3 : ---X---X---
\end{lstlisting}
\caption{Toffoli gate parallelism. Two parallel Toffoli gates can either be applied to; left) the same qubits acting as controls, or right) the target qubit and a shared control. Note: In the left diagram, the second Toffoli is not necessary and can be replaced with two CNOT gates, one before and another one after the first Toffoli gate.}
\label{fig:partof}
\end{figure}

In the following, we use a CCZ decomposition that maintains the Toffoli gate parallelism  when decomposed into Clifford+T. The decomposition is obtained after making the observation that two Toffoli gates are parallel whenever they are arranged like in Fig.~\ref{fig:partof}. There are two non-trivial situations: a) one wire is shared; b) two wires are shared. Whenever three wires are shared, the gates cancel each other.

Another practical observation is that, whenever two Toffoli gates are parallel, these can be formulated as CCZ gates which share one or two controls out of the three.

\begin{figure}[h]
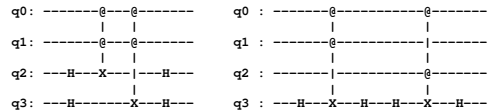

\linespread{.5}
\begin{lstlisting}[basicstyle=\ttfamily\bfseries\tiny]
          q0: -------@---@-------     q0 : -------@-----------@-------
                     |   |                        |           |
          q1: -------@---@-------     q1 : -------@-----------|-------
                     |   |                        |           |
          q2: ---H---X---|---H---     q2 : -------|-----------@-------
                         |                        |           |
          q3: ---H-------X---H---     q3 : ---H---X---H---H---X---H---
\end{lstlisting}
\caption{CCZ gate parallelism is whenever at least one control is shared. In this figure two wires are shared. The two Hadamards in the center of the right-most circuit will cancel.}
\label{fig:parccz}
\end{figure}

\begin{figure}[h]
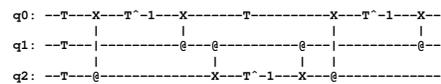

\linespread{.5}
\begin{lstlisting}[basicstyle=\ttfamily\bfseries\tiny]
                  q0: --T---X---T^-1---X-------T----------X---T^-1---X--
                            |          |                  |          |
                  q1: --T---|----------@---@----------@---|----------@--
                            |              |          |   |
                  q2: --T---@--------------X---T^-1---X---@-------------
\end{lstlisting}
\caption{CCZ gate Clifford+T decomposition that maintains Toffoli gate parallelism when a single wire is shared. The $\ket{q1}$ wire can be shared by two parallel Toffoli gates.}
\label{fig:decomp}
\end{figure}

In order to implement parallel Clifford+T decompositions of Toffoli gates, there has to exist a Clifford+T parallelisable decomposition of CCZ. The decomposition in Fig.~\ref{fig:decomp} can be used whenever two CCZ/Toffoli gates share a single wire. It can be noticed that the wire $\ket{q1}$ acts only as a control for the CNOTs, which are applied to the wires $\ket{q0}$ and $\ket{q2}$, and the T gate commutes with the control of the CNOTs. 

\begin{figure*}
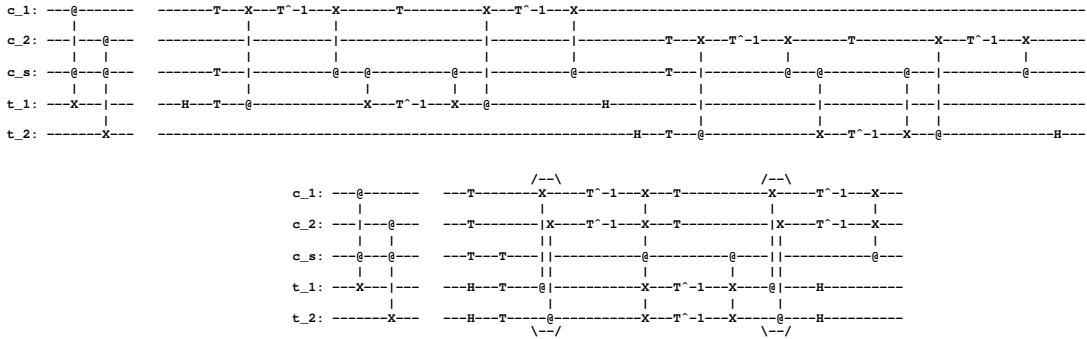

\linespread{.5}
\begin{lstlisting}[basicstyle=\ttfamily\bfseries\tiny]
                  c_1: ---@-------   -------T---X---T^-1---X-------T----------X---T^-1---X----------------------------------------------------------------
                          |                     |          |                  |          |
                  c_2: ---|---@---   -----------|----------|------------------|----------|-----------T---X---T^-1---X-------T----------X---T^-1---X-------
                          |   |                 |          |                  |          |               |          |                  |          |
                  c_s: ---@---@---   -------T---|----------@---@----------@---|----------@-----------T---|----------@---@----------@---|----------@-------
                          |   |                 |              |          |   |                          |              |          |   |
                  t_1: ---X---|---   ---H---T---@--------------X---T^-1---X---@--------------H-----------|--------------|----------|---|------------------
                              |                                                                          |              |          |   |
                  t_2: -------X---   ------------------------------------------------------------H---T---@--------------X---T^-1---X---@--------------H---
\end{lstlisting}
\begin{lstlisting}[basicstyle=\ttfamily\bfseries\tiny]
                                                                                    /--\                         /--\
                                                      c_1: ---@-------   ---T--------X-----T^-1---X---T-----------X-----T^-1---X---
                                                              |                      |            |               |            |
                                                      c_2: ---|---@---   ---T--------|X----T^-1---X---T-----------|X----T^-1---X---
                                                              |   |                  ||           |               ||           |
                                                      c_s: ---@---@---   ---T---T----||-----------@----------@----||-----------@---
                                                              |   |                  ||           |          |    ||
                                                      t_1: ---X---|---   ---H---T----@|-----------X---T^-1---X----@|----H----------
                                                                  |                   |           |          |     |
                                                      t_2: -------X---   ---H---T-----@-----------X---T^-1---X-----@----H----------
                                                                                    \--/                         \--/
\end{lstlisting}
\caption{Two Toffoli gates sharing a control wire. Parallelisation with canonical Toffoli decomposition.}
\label{fig:sharecontrol}
\end{figure*}

The wire ordering in Fig.~\ref{fig:decomp} does not play any role, because the output will still reflect the $(-1)^{xyz}$ phase flip. Effectively, a Toffoli gate can be obtained by surrounding the CCZ with two Hadamards on any of the wires. Advantageous configurations are whenever the Hadamards are placed such that the wire corresponding to $\ket{q1}$ is shared. The presented technique is similar to template based quantum circuit optimisation \cite{maslov2005quantum}, in the sense that the most advantageous Toffoli rewrite rule is chosen from the three possible decompositions based on the expected gate parallelism. During the writing of the manuscript, we found out that the decomposition from 
Fig. \ref{fig:decomp} has been also presented in \cite{amy2014polynomial}, but its effect on the parallelisation of Toffoli/CCZ gates has not been described in the subsequent literature.

\subsection{FANOUT: Shared Control}

Whenever two Toffoli gates share a control, the CCZ gate is decomposed such that the shared wire is the one that enables parallelism (i.e. $\ket{q1}$ in Fig.~\ref{fig:decomp}). This scenario appears in the FANOUT region of the QRAM.

\subsection{QUERY: Shared Target}

The QUERY is formed by a sequence of Toffoli gates conditioned by distinct pairs of $(b_i, m_i)$ wires. The only wire shared is the target. Therefore, the decomposition from Fig.~\ref{fig:decomp} can be used, but by making $\ket{q1}$ correspond to the target. For two consecutive Toffolis the H gates on the target wire will cancel, as illustrated in Fig.~\ref{fig:sharecontrol}. T-count optimisation is a side-effect of this kind of parallelism because along the shared target wire the T gates can pairwise be transformed into S gates.

\begin{figure*}
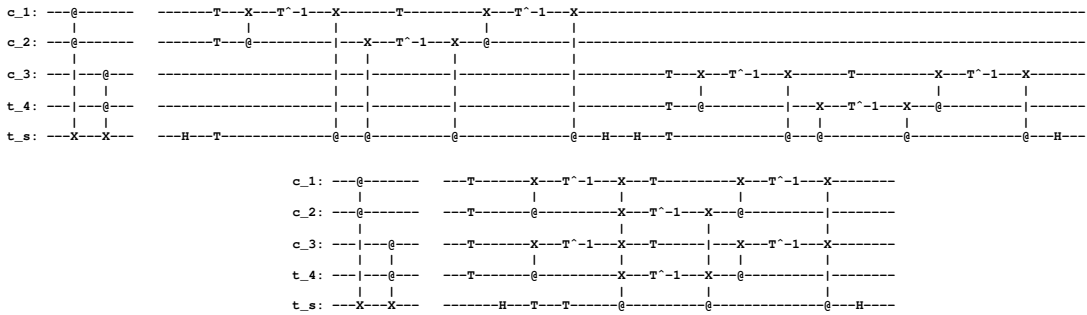

\linespread{.5}
\begin{lstlisting}[basicstyle=\ttfamily\bfseries\tiny]
                  c_1: ---@-------   -------T---X---T^-1---X-------T----------X---T^-1---X----------------------------------------------------------------
                          |                     |          |                  |          |
                  c_2: ---@-------   -------T---@----------|---X---T^-1---X---@----------|----------------------------------------------------------------
                          |                                |   |          |              |
                  c_3: ---|---@---   ----------------------|---|----------|--------------|-----------T---X---T^-1---X-------T----------X---T^-1---X-------
                          |   |                            |   |          |              |               |          |                  |          |
                  t_4: ---|---@---   ----------------------|---|----------|--------------|-----------T---@----------|---X---T^-1---X---@----------|-------
                          |   |                            |   |          |              |                          |   |          |              |
                  t_s: ---X---X---   ---H---T--------------@---@----------@--------------@---H---H---T--------------@---@----------@--------------@---H---
\end{lstlisting}
\begin{lstlisting}[basicstyle=\ttfamily\bfseries\tiny]
                                                      c_1: ---@-------   ---T-------X---T^-1---X---T----------X---T^-1---X--------
                                                              |                     |          |              |          |
                                                      c_2: ---@-------   ---T-------@----------X---T^-1---X---@----------|--------
                                                              |                                |          |              |
                                                      c_3: ---|---@---   ---T-------X---T^-1---X---T------|---X---T^-1---X--------
                                                              |   |                 |          |          |   |          |   
                                                      t_4: ---|---@---   ---T-------@----------X---T^-1---X---@----------|--------
                                                              |   |                            |          |              |   
                                                      t_s: ---X---X---   -------H---T---T------@----------@--------------@---H----
\end{lstlisting}
\caption{Two Toffoli gates sharing a target wire. Two of the T gates on the target wire can be reduced to an S gate. Simplification is not illustrated. Furthermore, half of the column of T gates at the beginning of the circuit could be eliminated in some circumstances (see Sec.~\ref{sec:total}).}
\label{fig:sharetarget}
\end{figure*}

\subsection{FANOUT: Compute Logical AND}

\begin{figure}[t]
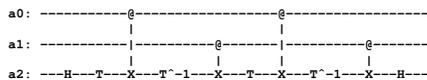

\linespread{.5}
\begin{lstlisting}[basicstyle=\ttfamily\bfseries\tiny]
             a0: -----------@------------------@------------------
                            |                  |
             a1: -----------|----------@-------|----------@-------
                            |          |       |          |
             a2: ---H---T---X---T^-1---X---T---X---T^-1---X---H---
\end{lstlisting}
\caption{The parallelisable logical AND has two wires acting always as control. This circuit is similar to the one from \cite{babbush2018encoding}. The target wire cannot be shared for parallelisation. If the qubit $\ket{a2}$ is initialised into $\ket{0}$, at the end of this computation will be in the state $(-i)^{a_0a_1}\ket{a_0a_1}$ which represents up to a phase shift the correct value of the Boolean AND operation between $a_0$ and $a_1$.}
\label{fig:logand}
\end{figure}

Approximate gates have been discussed since \cite{barenco1995elementary}, but their relevance for quantum circuit design was recognised once T-count optimisation became urgent. This is the case for the approximate CCZ/Toffoli which uses four T gates like in Fig.~\ref{fig:logand}.  The approximate Toffoli is also called a logical AND, when the target is an ancilla initialised to $\ket{0}$. However, after a logical AND the state is left with a phase shift, which has to be reversed, once the computed AND bit is not necessary anymore in the circuit (see following section).

\begin{figure*}[t!]
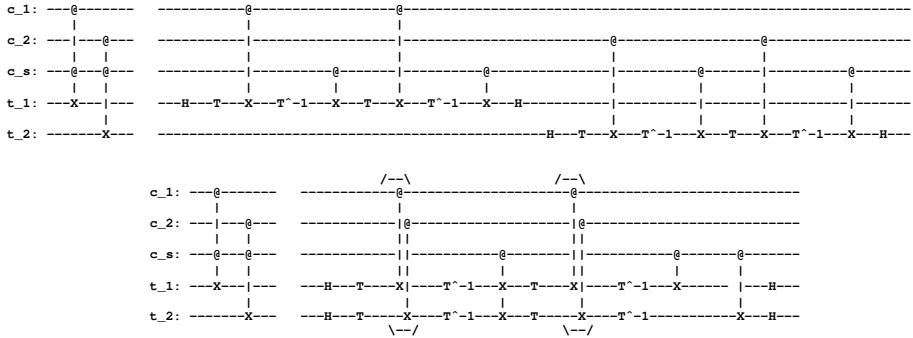

\linespread{.5}
\begin{lstlisting}[basicstyle=\ttfamily\bfseries\tiny]
                  c_1: ---@-------   -----------@------------------@----------------------------------------------------------------
                          |                     |                  |
                  c_2: ---|---@---   -----------|------------------|--------------------------@------------------@------------------
                          |   |                 |                  |                          |                  |
                  c_s: ---@---@---   -----------|----------@-------|----------@---------------|----------@-------|----------@-------
                          |   |                 |          |       |          |               |          |       |          |
                  t_1: ---X---|---   ---H---T---X---T^-1---X---T---X---T^-1---X---H-----------|----------|-------|----------|-------
                              |                                                               |          |       |          |
                  t_2: -------X---   -------------------------------------------------H---T---X---T^-1---X---T---X---T^-1---X---H---
\end{lstlisting}
\begin{lstlisting}[basicstyle=\ttfamily\bfseries\tiny]
                                                                 /--\                  /--\
                                    c_1: ---@-------   ------------@---------------------@----------------------------
                                            |                      |                     |
                                    c_2: ---|---@---   ------------|@--------------------|@---------------------------
                                            |   |                  ||                    ||
                                    c_s: ---@---@---   ------------||-----------@--------||-----------@-------@-------
                                            |   |                  ||           |        ||           |       |
                                    t_1: ---X---|---   ---H---T----X|----T^-1---X---T----X|----T^-1---X------ |---H---
                                                |                   |           |         |                   |
                                    t_2: -------X---   ---H---T-----X----T^-1---X---T-----X----T^-1-----------X---H---
                                                                  \--/                  \--/
\end{lstlisting}
\caption{Two Toffoli gates sharing a control wire. Parallelisation with logical AND versions of the Toffoli gate.}
\label{fig:andcompute}
\end{figure*}

Due to their lower T-count it is preferable, whenever possible, to use logical AND gates instead of more general CCZ/Toffoli gates. Two logical ANDs sharing a control can be decomposed like in Fig.~\ref{fig:andcompute}.

\subsection{FANIN: Uncompute Logical AND}

The inverse of the logical AND is its uncomputation. Because the $b_i$ qubits are ancillary, their usage is not necessary after the queries have been executed. Thus, these can be measured and a correctional CZ gate can be applied conditionally on the measurement result.

\subsection{The Parallel Bucket Brigade QRAM}
\label{sec:total}

The parallelised QRAM circuit is obtained by concatenating the parallelised circuits for FANOUT, QUERY and FANIN (e.g. Fig.~\ref{fig:complete}). One of the surprising results is that the depth of the FANOUT is  constant with respect to $q$ (cf. Eq.~\ref{eq:d}): the depth is reduced from $\mathcal{O}(Q\log{Q})$ where $Q=2^q$ to just $\mathcal{O}(\log{Q})$. This indicates that there is a significant speedup in computing $Q$ sums of the form $b_j=\sum_{i=0}^q (a_i\cdot 2^i)$ where $a_i$ are the bits of the address state vectors used as input to the QRAM (even in superposition).

Another interesting observation is that the T-count $T_{bbp}$ can be reduced with additional $2^n$ T gates if, after applying the QUERY, the memory is not entangled to the rest of the computation. In that case, the T gates on the $m_i$ wires will affect only the global phase of the $m_i$ states. This can be seen if the decomposition from Fig.~\ref{fig:decomp} is used in reversed gate order, such that, for example, in Fig.~\ref{fig:sharetarget} half of the T gates applied to $c_i$ in the leftmost column appear at the end of the circuit. Thus, by using parallel Toffoli gate decompositions one could obtain a T-count comparable to logical AND formulations. However, the major disadvantage of logical ANDs is their sequential nature, and being uncomputed through measurements.

\begin{figure}
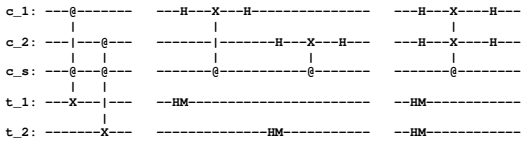

\linespread{.5}
\begin{lstlisting}[basicstyle=\ttfamily\bfseries\tiny]
         c_1: ---@-------   ---H---X---H---------------   ---H---X----H---
                 |                 |                             |
         c_2: ---|---@---   -------|-------H---X---H---   ---H---X----H---
                 |   |             |           |                 |
         c_s: ---@---@---   -------@-----------@-------   -------@--------
                 |   |
         t_1: ---X---|---   --HM-----------------------   --HM------------
                     |
         t_2: -------X---   --------------HM-----------   --HM------------
\end{lstlisting}
\caption{Logical AND uncompute. HM represents the measurement in the X basis (a Hadamard gate followed by a measurement in the computational basis). Depending on the measurement result a CZ gate correction is applied (here shown always and not conditioned on measurement result). The CZ gate is decomposed with the target being on the wire that is not shared. All measurements can be parallelised and the correction can be applied depending on the individual measurement results. In this figure, a parallel CNOT is applied for the case when both corrections would be required.}
\label{fig:anduncompute}
\end{figure}

\section{Discussion: Practical Speedups}
\label{sec:practical}

The presented exponential speedups in Sec. \ref{sec:theoretical_speedup} are theoretical. In practice, quantum circuits need to be mapped to hardware, or be compiled to error-corrected structures. The underlying physical architecture determines the achievable speedup which can be less than the ideally observed (e.g. theoretical) quantum gate level parallelism. Consequently, the available level of parallelism in a circuit may not result in exactly the same execution speedup. 

A concern could be that the physical realisations of QRAM will not be able to achieve an exponential speedup using the CNOT parallelisation scheme we presented. The question is if it is practically feasible to parallelise a single control multiple target CNOT, called parallel CNOT in Sec.~\ref{sec:par}, while maintaining  the speedup in practice. 

Similar to the theoretical speedup, the practical one is exponential as well. However, the speedup is scaled by some  factor (e.g. overhead). Independent of the design choices  of the QRAM circuit per se (main focus of this paper), the physical realisation (e.g. implementation which is not the main focus of this paper) of the QRAM in practice requires the usage of additional ancillae. Although we focus on the theoretical speedup, we demonstrate that CNOT parallelism is practically feasible by considering two perspectives: 1) fault-tolerant QRAM protected by the surface code. 2) un-error corrected QRAM;

\subsection{Fault-tolerant QRAM Speedup}

This work and the herein presented QUERY parallelisation was originally formulated with the surface code in mind. The surface code (in all its known variants and implementations) supports parallel CNOTs (Sec.~\ref{sec:par}). The achieved QUERY speedups are $\mathcal{O}(1)$ in the case of surface code protected quantum circuits. Moreover, CNOT logical gates are transversal in the surface code, and the classical post-processing time of these logical gates is constant (see  \cite{raussendorf2003measurement}).

The physical realisation of the surface code requires nearest neighbour physical connectivity. Mapping surface code circuits to a two-dimensional lattice or 3D cluster state \cite{paler2014mapping} is an efficient procedure both in terms of resulting number of gates (from logical to physical implementation - the number of qubits and gates to implement a logical operation is approximated by a polynomial function), as well as algorithmic (the mapping procedure has polynomial complexity).

The remaining concern could be related to the arrangements of the surface code protected qubits such that the exponential parallel CNOT complexity is achieved. However, this can be solved without introducing any overhead by ordering the logical qubits in pairs, similarly to the one presented in \cite{arunachalam2015robustness}: 
$a_0,\ldots,a_q,\ldots b_i,m_i,b_{i+1},m_{i+1}\ldots target$

With respect to the surface code implementation of the bucket brigade QRAM, it should be noted that QRAM query time is exponentially shorter after exponentially reducing the depth of the circuits. The estimations in \cite{di2019fault} mention 0.35 ms for querying a 4KB QRAM (15 bit addresses) that is protected by the surface code. The presented optimisation \emph{would} reduce the query time of the surface code error corrected QRAM by a factor of $2^{15}$ to approximately 11 ns.

\subsection{Un-Error Corrected QRAM Speedup}

If the QRAM circuit would be executed un-error-corrected directly on a quantum chip, the connectivity of the chip determines the exact achievable speedup. However, the practical speedup would still be exponential, by making the strong assumption that \emph{ sufficient ancillae are available}.

The theoretical exponential speedups are scaled only by a factor on the order of $q$, which is the number of address qubits (see Sec.~\ref{sec:intro}). In particular, for a given address length $q$, the value of $q$ is a \emph{constant}: once the memory size is known and configured for an algorithm, this cannot be changed anymore.

A parallel CNOT can be implemented using a binary tree where the CNOT control is the root, and the CNOT targets are the leaves. The new ancillae are initialised in $\ket{+}$ and used to construct in $\log(2^q) = q$ steps a large GHZ (Greenberger–Horne–Zeilinger) state to encode the state of the control qubit. Afterwards, the parallel CNOT is a transversal application of CNOTs between the GHZ state and the original targets implement. This naive construction illustrates that theoretical $O(1)$ speedups: 1) are delayed by a factor of $q$ which is the time necessary to construct the trees; 2) are not impacted by the number of additional ancillae, although the resource efficiency of this scheme is reduced.

It should be noted that the GHZ construction doubles the number of qubits in the QRAM, from $Q_{bbp}$ to $2\cdot Q_{bpp}$. It does not introduce an overhead of the form $2^{Q_{bpp}}$.

The naive tree construction is structurally very similar to the FANOUT region of the bucket brigade QRAM. The difference is that the FANOUT region is performing logical-AND operations (Toffoli gates) which are more complex than the logical-XOR operations necessary for the parallel CNOT.

It is possible to implement the parallel CNOT using less ancillae, compared to 
the naive construction, by using non-fault-tolerant cluster states \cite{raussendorf2003measurement}. The cluster state and measurement-based formalism are another proof that the exponential speedup is scaled by a constant factor, because ``gates from the Clifford group do not contribute to the complexity of a quantum algorithm'' \cite{raussendorf2003measurement}. The parallel CNOT is a Clifford gate.

\section{Conclusion}

Quite a few quantum algorithms require access to information stored in a database like manner. To this end, QRAMs were proposed in general, and the bucket brigade QRAM model in particular. 

A bucket brigade QRAM includes three stages: 1) FANOUT, where the input addresses are used to compute exponentially many memory pointers to all possible memory cells; 2) QUERY, where the memory cells are queried using Toffoli gates, and in this process the memory pointers are used to control the queries; 3) FANIN, where the memory pointers are uncomputed, leaving the ancillae wires in their original state before FANIN. The exponentially many memory pointer wires increase the speed of the addressing, but, as shown in this paper, have also another advantage: can be used to massively parallelise the queries when considering the Clifford+T gate set decomposition of the Toffoli gates.

We construct bucket brigade QRAM circuits having a constant QUERY depth. We achieve this by using advantageous Toffoli gate decompositions, and do not introduce any additional ancilla qubits into the QRAM circuit (except the already available memory pointers/control signal wires). The depth, when formulated with Clifford+T gates, is independent of the number of queries, and it depends solely on the number of bits necessary to address the memory. Compared to state of the art bucket brigade implementations, we reduced the depth of QUERY from $\mathcal{O}(2^q)$ to $\mathcal{O}(q)$ in the worst case when any of the $2^q$ memory cells QRAM are being queried.

Incidentally, the presented construction has a T-count more than half smaller, compared to the existing Clifford+T formulations. Our construction reduces significantly the depth of the topological assembly \cite{paler2017synthesis} representing the bucket brigade circuit and, thus, also the distance of the surface code to protect the assembly.

The parallel QUERY construction shows that, if quantum hardware would not be a scarce resource, exponential query speedups are possible compared to state-of-the art QROM designs (Sec.~\ref{sec:practical}). The QROM uses exponentially less wires than bucket brigade, but has exponentially slower querying in the worst case. The QROM circuits execute queries sequentially, and to the best of our knowledge no parallel construction seems possible.

\begin{figure}[t!]
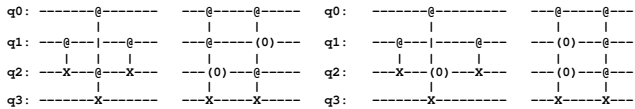

\linespread{.5}
\begin{lstlisting}[basicstyle=\ttfamily\bfseries\tiny]
q0: -------@-------   ---@-----@-----   q0:   -------@---------   ---@-----@---
           |             |     |                     |               |     |
q1: ---@---|---@---   ---@-----(0)---   q1:   ---@---|-----@---   ---(0)---@---
       |   |   |         |     |                 |   |     |         |     |
q2: ---X---@---X---   ---(0)---@-----   q2:   ---X---(0)---X---   ---(0)---@---
           |             |     |                     |               |     |
q3: -------X-------   ---X-----X-----   q3:   -------X---------   ---X-----X---
\end{lstlisting}
\caption{Parallelising operations by introducing Toffolis (\texttt{{(0)}} represents a negative control) reduces computational depth. In the worst case, where no other optimisations are available, it increases T-count. These circuit identities are inverse to the ones from \cite{rahman2014templates}.}
\label{fig:template}
\end{figure}

The parallelisation of bucket brigade was achieved without introducing any additional ancillae, and future work will parallelise quantum circuits using templates like the one presented in Fig.~\ref{fig:template}. It may seem counter intuitive, that it may be more resource efficient to include T gates while not introducing ancilla: in Fig.~\ref{fig:template} a single Toffoli is transformed into two Toffoli gates, thus the number of T gates increases from 7 to 14. However, the introduced T gates may cancel, due to efficient Toffoli gate parallelism in the overall circuit. We argue, that the cost of large scale quantum error-corrected quantum circuits is not necessarily related to T-counts and state distillations, but to the error-corrected associated with the error-corrected Clifford operations \cite{paler2019clifford}. This includes identity operations on unused wires and ancillae, as for example in carry save adders.

\begin{acknowledgments}
A.P. was supported by a Google Faculty Research Award, and the NUQAT project funded by the University Transilvania Brasov. We thank Olivia Di Matteo for her very valuable feedback during the preparation of the circuits and the writing of the manuscript, and Simon Devitt for feedback on the final manuscript.
\end{acknowledgments}
\bibliography{__main}

\end{document}